# Reply to: Large-scale quantitative profiling of the Old English verse tradition

**Petr Plecháč**
Institute of Czech Literature, Czech Academy of Sciences, Czech Republic

**Andrew Cooper**
Stockholm University, Sweden

**Benjamin Nagy**
University of Adelaide, Australia

**Artjoms Šela**
University of Tartu, Estonia

In *Nature Human Behaviour* 3/2019, an article was published entitled *Large-scale quantitative profiling of the Old English verse tradition*[1] dealing with (besides other things) the question of the authorship of the Old English poem *Beowulf*. The authors provide various textual measurements that they claim present "serious obstacles to those who would advocate for composite authorship or scribal recomposition" (p. 565). In what follows we raise doubts about their methods and address serious errors in both their data and their code. We show that reliable stylometric methods actually identify significant stylistic heterogeneity in *Beowulf*. In what follows we discuss each method separately following the order of the original article.

## Sense-pause analysis

The first argument in favor of unitary authorship of *Beowulf* is based on the distribution of selected punctuation marks, namely on the value of what portion of these occur in the non-final position of the verse line. This metric is shown to yield similar values in texts supposedly written by the same author (Cynewulf's signed poems *Elene* and *Juliana*) and different values in texts on similar topics written by different authors (*Genesis A* vs. *Genesis B*; *Christ I* vs. *Christ II* vs. *Christ III*). Two different editions of *Beowulf* are reported not to exhibit significant differences in this respect before and after line 2300 (the point where most theories of composite authorship mark the divide).

Evidence against this method's ability to distinguish authorship is actually provided by the authors themselves when they observe that they have "found a marked difference in the intraline-to-total sense-pause ratio between *Genesis A* and *B* … sense-pause analysis can distinguish between passages of Old English verse about similar subject matter but composed by different poets" (p. 3). Consultation of their sample shows that the values actually measured are both from *Genesis A* (lines 1–234 and lines 852–2936 respectively). The radically metrically different *Genesis B* is not actually included in their data set, although the first line of *Genesis B* is appended to the first sample. The "marked difference" is thus found between two pieces of a single text, the unitary authorship of which has never been questioned.

We have also identified several errors in the code (enumerated in the Supplementary Information). Beside these errors let us also address more general issues.

- The authors adopted this method from Fitch[2], where the ratio was used to trace the chronology of dramatic texts *within* the works of one author (intra-line pauses functioned as a substitute of another feature: in-line speaker change in plays). It has not been demonstrated that this metric is valid either for authorship attribution, or for genres outside drama. Following Fitch's analysis, one would actually expect the ratio to be varied in the works of one author due to chronological changes.
- The association between punctuation and sense-pauses puts the authors at the mercy of earlier editors and their usually German punctuation traditions[3]. In addition, in some manuscripts the punctuation marks metrical boundaries rather than syntactic ones.
- The metric seems to be very unstable within a single text. When texts are partitioned into 100 lines long samples, their ratios vary substantially. There is no significant difference between samples from *Christ I* and *Christ II* ($t(6) = 0.94$; $P = 0.3838$), nor between *Christ I* and *Christ III* ($t(9) = 1.03$; $P = 0.3319$). The significant difference between samples from *Genesis A* and *B* ($t(27) = 2.07$; $P = 0.0483$) may be explained by their different line lengths (*A*: 9.72 syllables per line; *B:* 12.07 syllables per line), due to *B*'s continental origin.

## Metre

The authors examined the frequencies of half-line patterns using a simplified Sievers scansion. The basis of their claim, in this section, is that the patterns are used at a fairly constant rate, before and after line 2300 (half-line 4600). Statistically, they measured Pearson's *r* when comparing a running count of pattern usage ("ordinal index of metre incidence"), compared to the half-line index. Since both measurements are monotonically increasing, it is not surprising to find that they appear to be positively correlated. What Pearson's *r* does not really tell us is the *relative density* of the patterns in any given section of text. The authors offer no evidence that this metric is able to differentiate OE authors and in what follows we question its validity. For now, visualising the relative proportions as a stacked graph (Fig. 1) gives a much more realistic feel for the variation in half-line patterns as we move through the text.

A much larger concern, for this section, is the authors' decision to analyze half-line patterns instead of full-line patterns. The basic unit of OE poetry is, like most verse types, the line, not the half-line[4]. We analysed the relationship between the two halves of the line, and found that they are not statistically independent, suggesting that such an analysis is better performed on full lines. When using all 25 possible line patterns, the variation is much clearer.[1] Fig. 2 allows for the claim that a change can be seen around line 2300, however other comparable peaks are also found in earlier sections.

---

1  It would be interesting, here, to consider the wave-like patterns of metrical variation with the action of the poem in the corresponding sections. Are certain patterns connected to certain kinds of action, or topics?

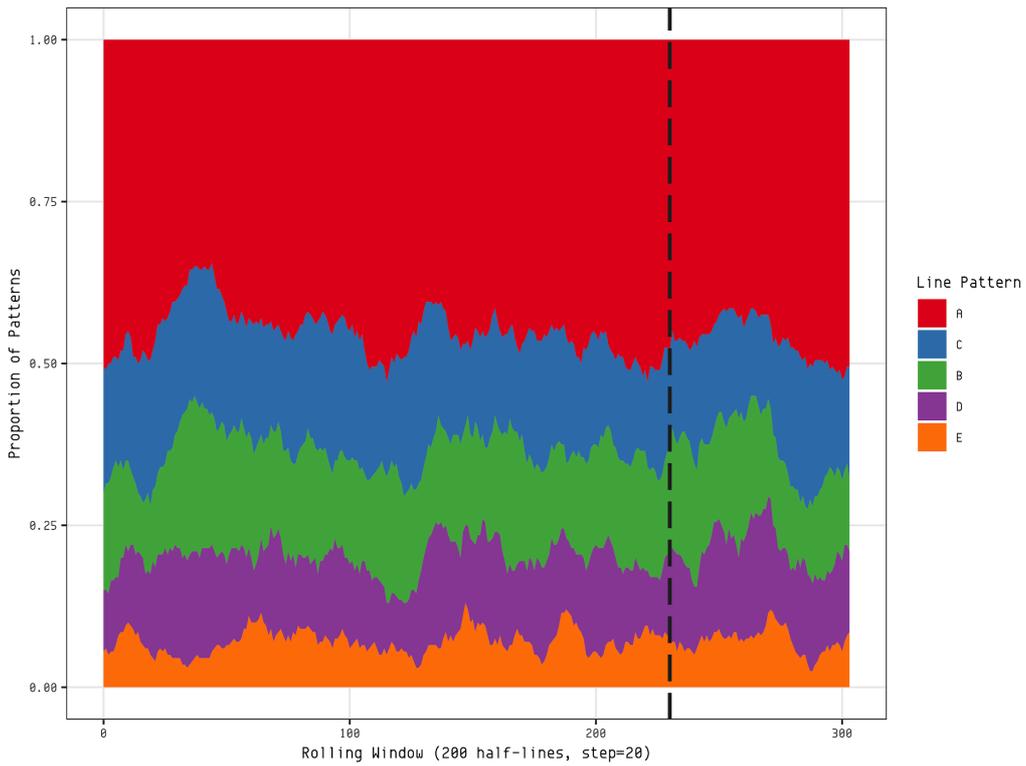

Fig. 1 | Half-line pattern variation. Relative proportions of each pattern, taken over a 200-line rolling window throughout the text. The vertical line shows the start of the disputed section.

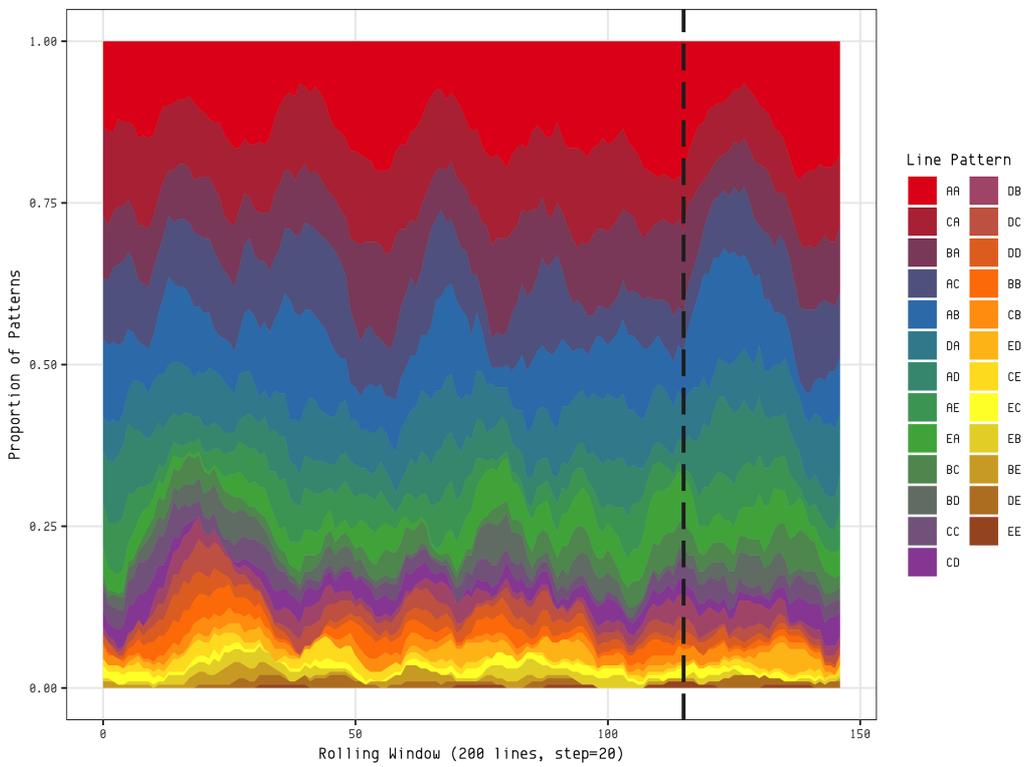

Fig. 2 | Full-line pattern variation. Relative proportions of each pattern, taken over a 200-line rolling window throughout the text. The vertical line shows the start of the disputed section.

Finally, we performed our own analysis of the pattern frequencies. The standard test, here, given that we are comparing two distributions, is a simple chi-square. There are two types of test that could be chosen, which are subtly different. The first is a "chi-square test for homogeneity" which tests the null hypothesis that both sections of the poem are samples from a single distribution. This is the more conservative test. The other approach is a "chi-square test for goodness of fit", which tests the null hypothesis that the distribution of patterns in the second half of the poem matches the distribution in the first half. We performed these tests on the half-line patterns as well as the full-line patterns. In addition, we empirically verified the full-line tests using a bootstrap simulation.

As can be seen from Table 1, the results for half-line patterns are inconclusive. However, both of the tests on full-line patterns are informative. In particular, the full-line goodness-of-fit test result suggests that there is statistically significant evidence for *different* authorship. It is not our intent, in this paper, to attempt to categorically prove multiple authorship, but this must be certainly be considered a serious blow to the claim of single authorship.[2]

| Line Type | Test | p-value |
| --- | --- | --- |
| Half | Homogeneity | 0.4649 |
| Half | Goodness of Fit | 0.303 |
| Full | Homogeneity | 0.121 |
| Full | Goodness of Fit | 0.0112 |

Table 1: Chi-Square test results, performed on half lines and full lines, comparing the pattern distributions before and after line 2300. The result from the goodness-of-fit test using full-line patterns is statistically significant.

## Hapax compounds

Hapax compound analysis methodologically follows the analysis of metre. Pearson's *r* is measured for a running count of hapax compounds (those occurring only once in the entire corpus) compared to the line index. Divergences in slope are considered to be an indicator of shift of authorship.

We were able to reproduce the results reported in Supplementary Fig. 4 in the original article, which claims that hapax compound analysis is suitable for determining authorship, except for Supplementary Fig. 4b, where the linear fit for 3 random partitions of *Exodus* is shown to have similar slopes. In our results when *Exodus* is split into 3 samples of the same length, the slope differs significantly for the middle sample (Supplementary Fig. 2a). In general, this section suffers from the same questionable assumption as above – that the consistency of the slope for count data is a stylometric indicator. This fragility of this method would be easily demonstrated with adversarial data, but it is also clear from the data used in the original study. When e.g. *Elene* is divided into halves, the difference between the slopes (Supplementary Fig. 2b) is comparable to that reported between *Genesis A* and *B* (Supplementary Fig. 4a of the original article). Even more strikingly, we then merged three texts for which no common authorship can be adduced: *Elene, Genesis B*, and *Phoenix*. The slopes match, with a correlation coefficient comparable to that claimed by the authors for *Beowulf*. The same can be see when merging *Genesis A* and *Andreas* (Supplementary Fig. 2c). All this questions the ability of this method to distinguish authorship and, hence, the relevance of the result reported for *Beowulf*.

---

2   We performed our tests on the half-line data as provided by the paper authors, and produced our full-line data by taking sequential pairs. However, we must note that some half-lines are missing from the source data, which means that some of our full-lines are out of alignment (we have the end of one line paired with the beginning of the next). This does not seem to be a major statistical problem when comparing sections to each other, but it means that the relative frequencies of the full-line patterns will be incorrect. If future work were to attempt to claim heterogeneity, it would probably be advisable to use verified scansion data based on full lines.

## Shared compounds

Additional support for the unitary authorship of *Beowulf* is sought in the analysis of shared compounds. The authors compare the number of compounds that are shared by a particular pair of poems to the model where all the compounds are randomly distributed across the corpus. They find a strong correlation between all Cynewulf poems as well as between *Beowulf* lines 1–2300 and *Beowulf* lines 2301–end.

We were unable to replicate the results summarized in Fig. 3 of the article. In our tests, we obtained values which differ to such an extent that it cannot simply be due to random variation in the models. This was double-checked using both the code provided by authors and our own implementation. In our results both parts of *Beowulf* are much less correlated than indicated by the original figure.

In addition, the results for the original dataset say nothing about how the method behaves in cases when different authorship comes within a poem on a single topic (which is a null hypothesis for *Beowulf*). Therefore we have measured the correlation also for *Christ I, II, III*. As can be seen in Fig. 3, despite the fact that these three poems are claimed to be written by different authors, their mutual correlations do not significantly differ from those of poems signed by Cynewulf ($t(7) = 0.19$, $P = 0.8567$).[3] This suggests that the method is sensitive to a common topic and as such can not be accepted as "further support for unitary composition" of *Beowulf*.

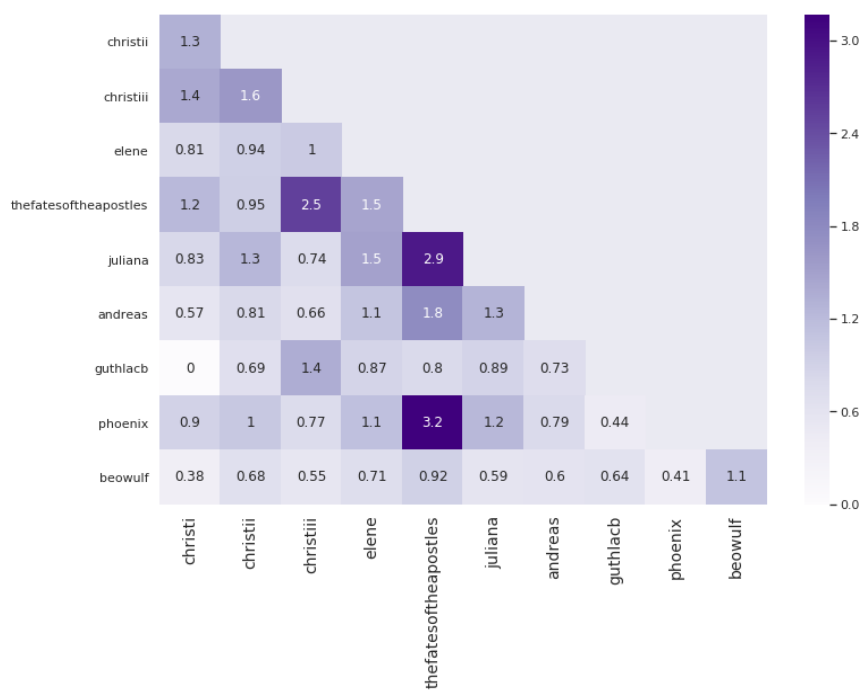

Fig. 3 | Shared compounds. Replication of original Fig. 3 with *Christ I* and *Christ III* added to the chart.

---

[3] The strong correlation between the Fates of the Apostles and almost all the other poems seems to be due to the low number of compounds found in it and to the fact that each of them occurs once (its compounds type-token ratio is significantly higher than for the other poems). The random model thus assigns a very low probability of compounds being shared with other texts.

### N-gram frequencies

The authors finally claim that the result of cluster analysis of poems based on 25 most frequent character trigrams, which is summarized in Fig. 4 of the article (p. 5), presents further evidence for unitary authorship ("in line with our other studies, *Beowulf* 1–2300 and 2301–end cluster together", p. 5). This is not valid unless there are other texts in the corpus written by an author to which would null hypothesis assign one of *Beowulf*'s parts. The only thing such a dendrogram says about *Beowulf* is that its two parts are stylistically more similar to each other than to other texts in the corpus written by different people on a different topic. Also, to analyze only 25 most frequent *n*-grams is a low number considering the number of tokens in the corpus.

To scrutinize *Beowulf*'s inner stylistic relations and to evaluate the method at the same time, we took the six longest texts in the corpus (*Beowulf; Genesis A+B; Andreas; Christ I+II+III; Guthlac A+B; Elene*) and divided them into 300-line samples. We performed the cluster analyses (cosine distance with complete linkage) based on frequencies of the 500 most frequent 3-grams and also the 25 most frequent 3-grams, to evaluate the original design. The results clearly indicated that the former approach yields high accuracy in distinguishing the authorship and that it actually suggests significant stylistic heterogeneity in *Beowulf*. On the other hand, the accuracy of the latter approach was significantly lower (see Supplementary Fig. 3 and 4).

To be able to verify more precisely the method's ability to trace the shifts in authorship within a single text, we repeated the analysis with the 500 most frequent 3-grams but divided each text into overlapping samples. We retained the frame at 300 lines but reduced the step to 100 (thus sample 1 of each text contained lines 1–300, sample 2 lines 101–400 etc.). The resulting dendrogram given in Fig. 4 indicates that the method is very reliable:

- Samples from each text cluster more closely together than with samples from other texts, except for *Elene* and *Andreas* (cf. possible Cynewulfian authorship of both). This is also true for the cluster containing samples primarily from *Christ I* and *Christ II*, which appears closer to the cluster containing samples from *Guthlac* than to the cluster containing samples from *Christ III*.
- Samples coming from or consisting mostly of the lines from *Christ I* are clustered separately from samples coming from or consisting mostly of the lines from *Christ II*. These are then separated from samples coming from *Christ III* or containing mixed content from *Christ II* and *Christ III*.
- Samples coming from *Guthlac A* are clustered separately from samples coming from *Guthlac B* or containing mixed content from *Guthlac A* and *Guthlac B*.
- Samples coming from *Genesis A* are clustered separately from samples coming from *Genesis B* or containing mixed content from *Genesis A* and *B*.

*Beowulf* splits into two distant clusters. (Their distance is actually greater than the distance between *Christ I* and *Christ II* or *Guthlac A* and *B*.) This contrary to the conclusion of the original article, and suggests that there is a lot of inner stylistic heterogeneity.[4] Such clustering of *Beowulf* was found to be very robust against the changes in both the number of *n*-grams analyzed (provided it is kept reasonably high) and the *n*-gram length (tested with 100, 150, 200, ... , 1000 most frequent 2-, 3-, 4- and 5-grams, cf. Supplementary Fig. 5).

---

4   The boundary however seems not to be located in the neighborhood of line 2300 but rather closer to where the scribal hand changes in the manuscript (line 1939).

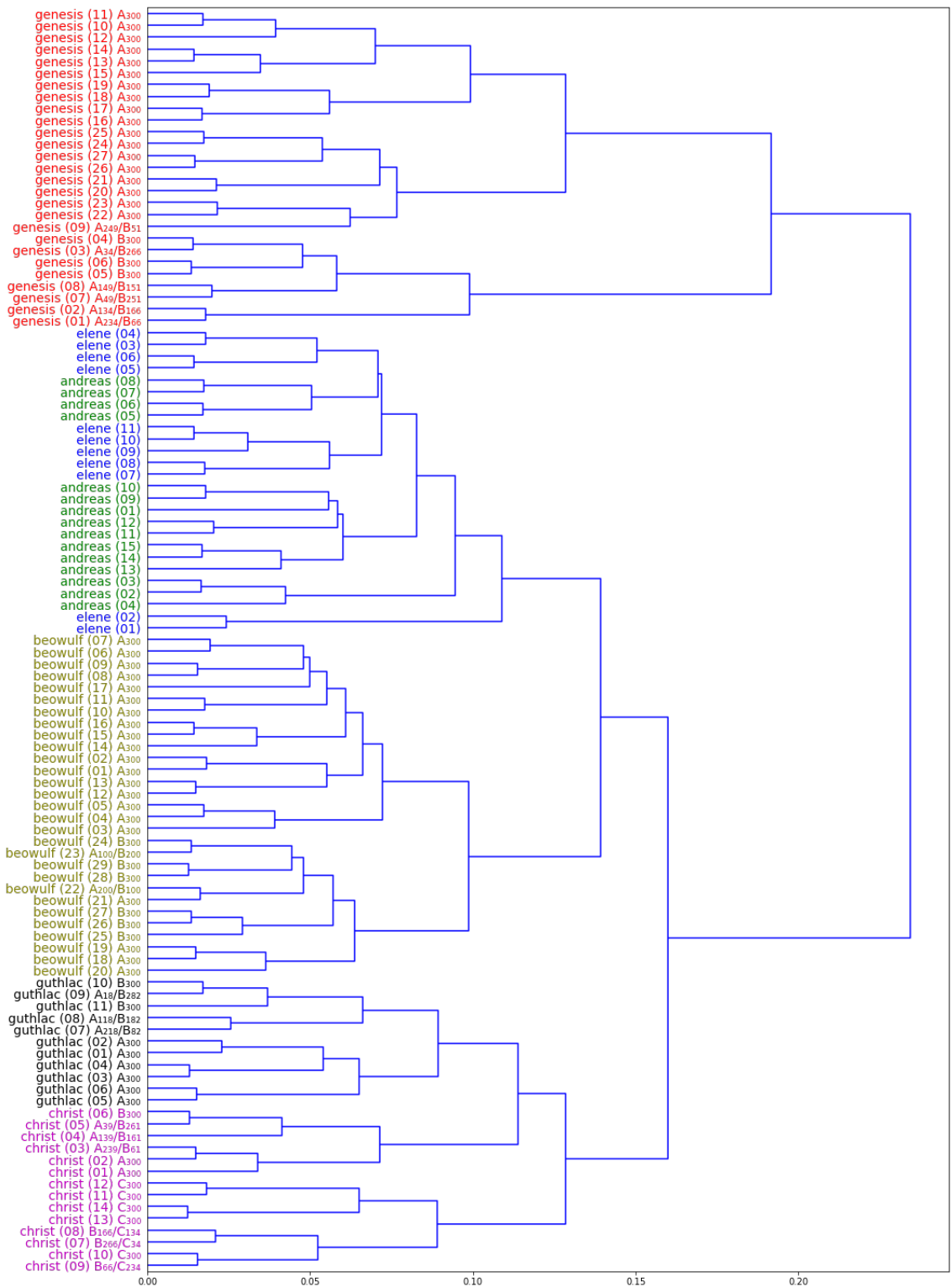

Fig. 4 | Dendrogram of overlapping samples of 300 lines, drawn from the 6 longest poems in the corpus based on the 25 most frequent character trigrams (cosine distance, complete linkage method). For *Beowulf*, *Christ*, *Guthlac* and *Genesis* the letter indicates the part from which the lines come; the subscript indicates the number of lines from a given sample.

## Conclusions

Our attempt to replicate this study shows that all four major features used by the authors for "quantitative profiling" of OE verse suffer from either undetermined methodological value (leading to misinterpretation of the authorship attribution results), suboptimal implementation, or both. We also found significant errors in code, sampling problems and unexplained omissions of data points. In addition, one part of the reported results could not be reproduced. All of this calls into question the strength of the major claims of the paper - including the unitary authorship of *Beowulf*.

There is one more general note to draw from our conclusions: modern computational stylometry spends the majority of time not in solving authorship attribution mysteries, but in carefully evaluating and testing methods and features under controlled conditions[5,6]. This is necessary to understand the discriminatory power of each technique or feature and to maximize contextual information when drawing conclusions. The paper in question reminds us of the importance of this methodological caution.

## Data availability

The data necessary to reproduce the analyses are provided at

https://github.com/versotym/beowulf.

## Code availability

The complete set of analysis code is available at https://github.com/versotym/beowulf, allowing all the analyses to be fully examined.

# Reply to: Large-scale quantitative profiling of the Old English verse tradition

## *Supplementary information*

**Petr Plecháč**
Institute of Czech Literature, Czech Academy of Sciences, Czech Republic

**Andrew Cooper**
Stockholm University, Sweden

**Benjamin Nagy**
University of Adelaide, Australia

**Artjoms Šela**
University of Tartu, Estonia

List of errors in the code calculating intraline-to-total sense-pause ratio

- In the article sense-pauses are defined as "breaks in speech typically denoted by any punctuation mark other than a comma" (p. 2), in the *Methods* section these marks are enumerated as . ? ! ; : ( ) - ' ' " and ". The original script used to calculate the ratios ([https://github.com/qcrit/NHB-2018-OEstylometry/blob/master/sense_pauses/tools.py](https://github.com/qcrit/NHB-2018-OEstylometry/blob/master/sense_pauses/tools.py), line 6) however identifies only the first seven of them, others are disregarded.

- The fact that commas are not taken into account but are not deleted from the texts prior to the analysis causes errors. For instance in the line *fœmnan lufian, (hine fyrwet bræc)*, the right bracket is incorrectly classified as an intra-line as it is followed by another character and not the end of the line directly.

- Finally dots are used in the corpus not only as the punctuation marks, but also as a part of ellipsis indicating missing pieces of texts (*e.g. þæt geond eorðb... ...g eall eagna gesihþe*). These cases (frequent namely in *Christ I*) are incorrectly classified as periods.

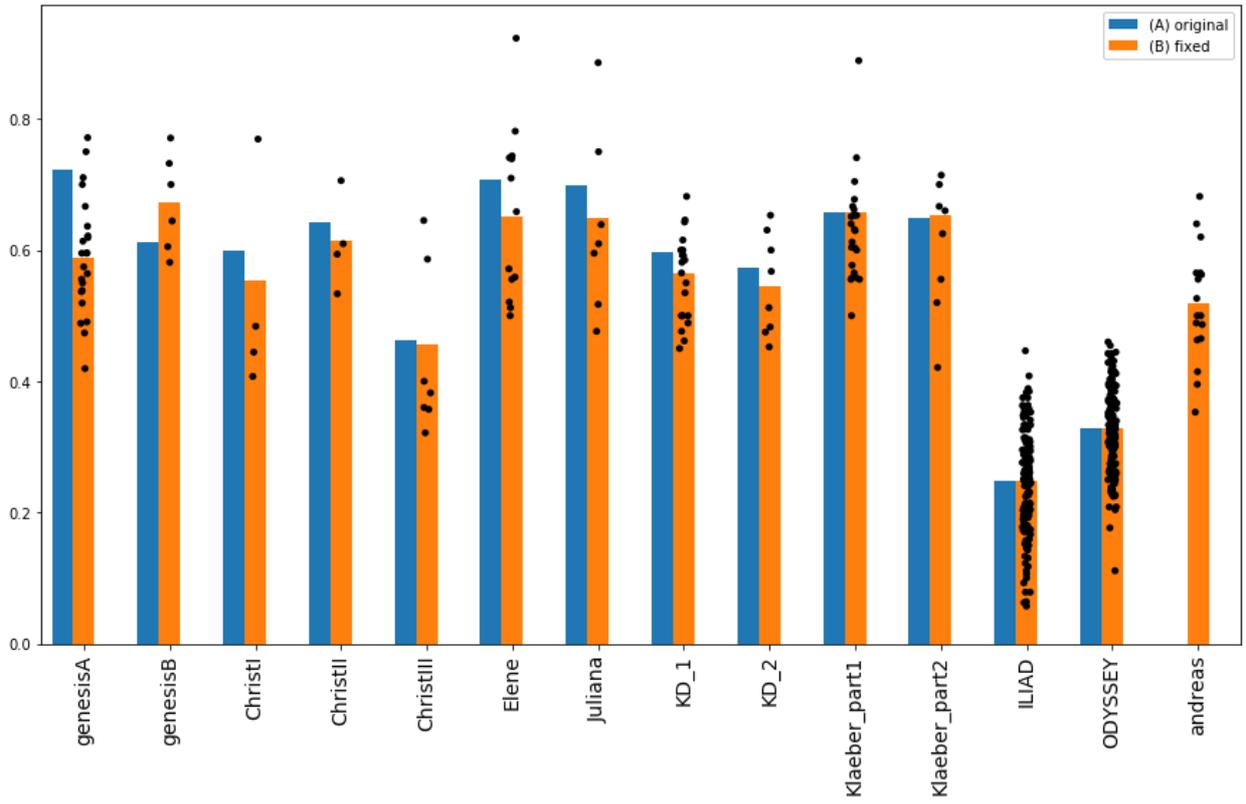

Supplementary Fig. 1 | Ratio of intraline to total sense-pauses. Replication of original Fig. 2A published in Neidorf. et al. against results when errors fixed. Dots represent ratios when text partitioned into 100 lines long samples.

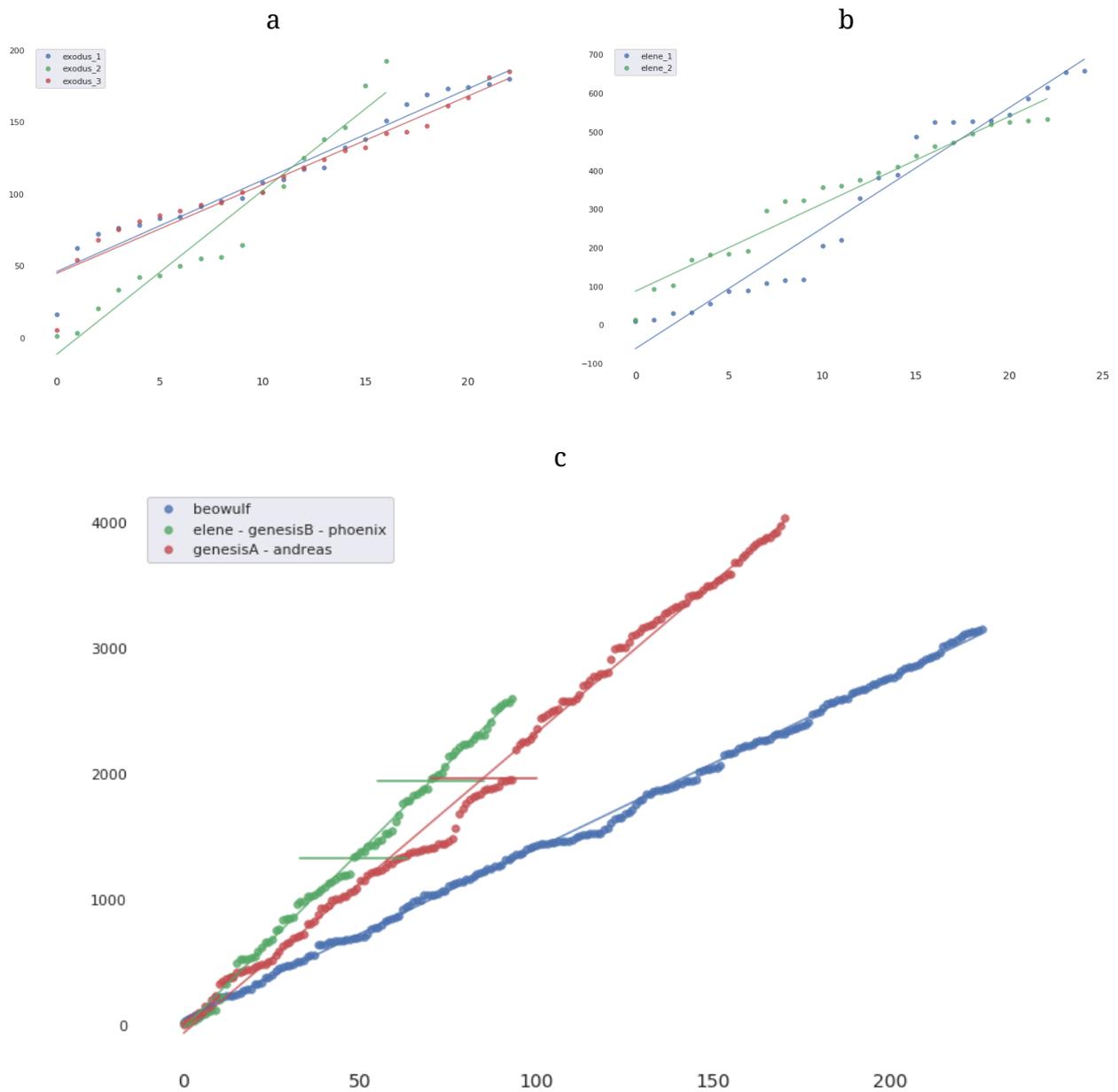

Supplementary Fig. 2 | Usage of hapax compounds. **a**. Slope differs significantly when *Exodus* partitioned into three samples of the same length (sample 1: slope = 6.36; $r$ = 0.9783; sample 2: slope = 11.38; $r$ = 0.9753; sample 3: slope = 6.17; $r$ = 0.9705). **b.** Difference between slopes for two samples of *Elene* is comparable to the difference reported between *Genesis A* and *B* in supplementary Fig. 4 of the original article (sample 1: slope = 31.25; $r$ = 0.9792; sample 2: slope = 22.67; $r$ = 0.9798). **c.** When *Elene, Genesis B* and *Phoenix* are merged together, the quality of linear fit is comparable to that of *Beowulf* (slope = 28.05; $r$ = 0.9983). Merging together *Genesis A* and *Andreas* produces a good quality fit as well (slope = 23.88; $r$ = 0.9972).

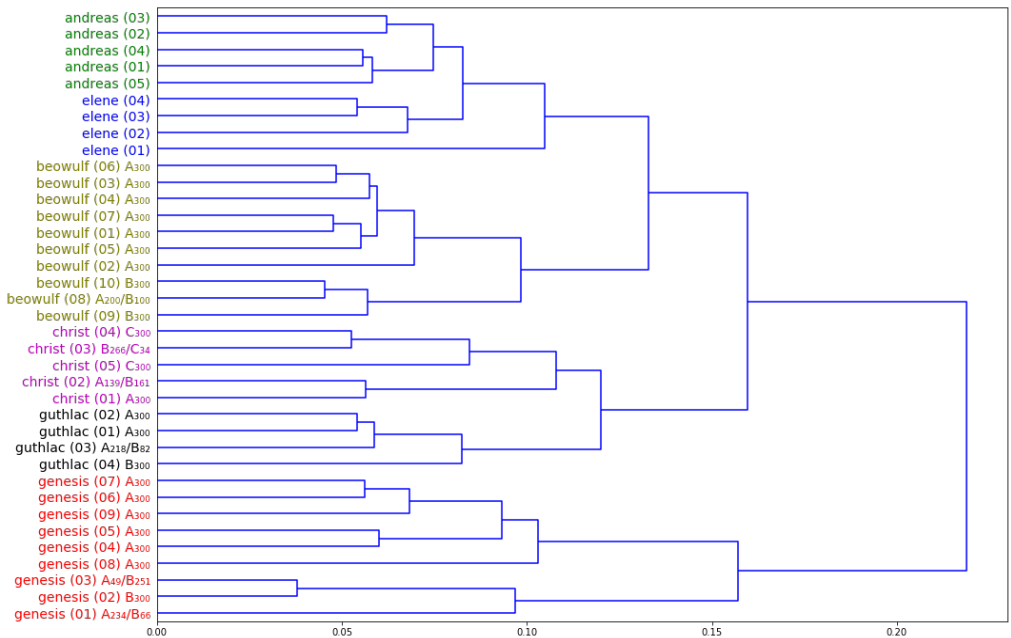

Supplementary Fig. 3 | Dendrogram of 300-lines samples drawn from the 6 longest poems in the corpus based on 500 most frequent character trigrams (cosine distance, complete linkage method). For *Beowulf, Christ, Guthlac* and *Genesis* the letter indiciates from which part does the lines come from, subscript indicates the number of lines from a given sample.

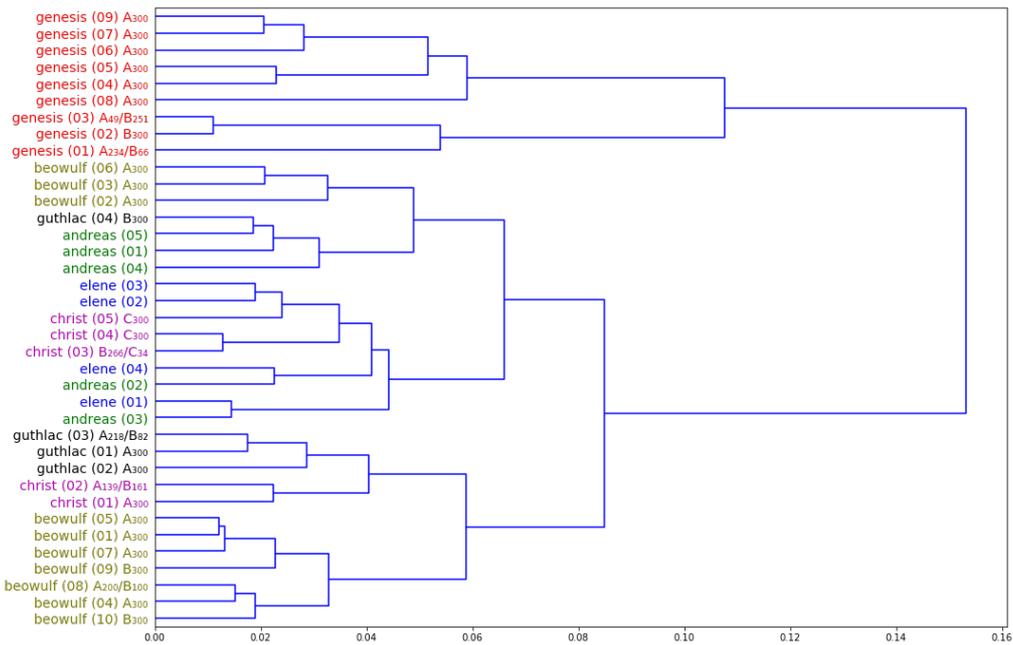

Supplementary Fig. 4 | Dendrogram of 300-lines samples drawn from the 6 longest poems in the corpus based on 25 most frequent character trigrams (cosine distance, complete linkage method). For *Beowulf, Christ, Guthlac* and *Genesis* the letter indiciates from which part does the lines come from, subscript indicates the number of lines from a given sample.

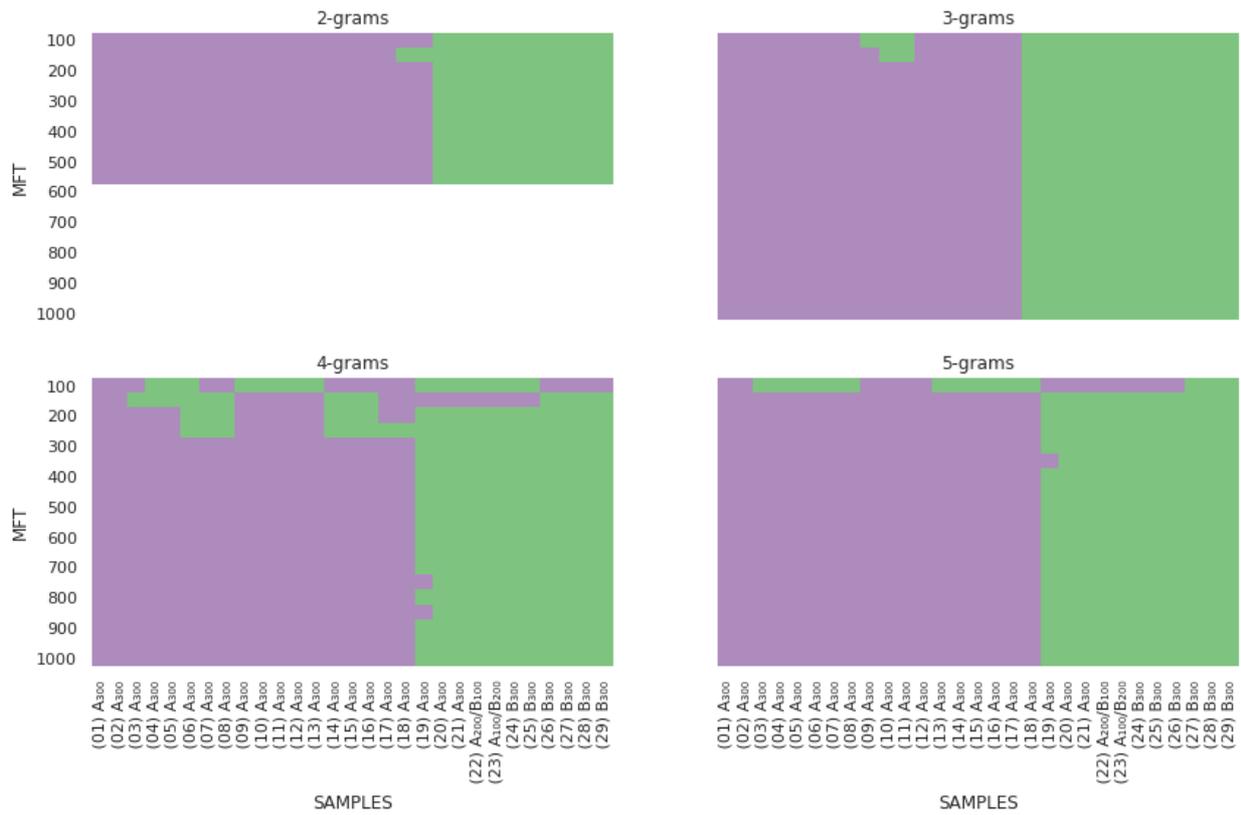

Supplementary Fig. 5 | Hierarchical agglomerative clustering of overlapping samples of 300 lines from *Beowulf* based on 100, 150, 200, ..., 1000 most frequent character 2-grams, 3-grams, 4-grams, and 5-grams. Color indicates to which of the top two clusters the sample belongs.